\documentclass[manuscript,screen,nonacm]{acmart} 
\setcopyright{cc} 
\acmConference[]{} 

\newcommand{\workshopname}{AI-XR for multimodal Human-AI Experience, JetBrains Research}
\newcommand\extrafootertext[1]{
    \bgroup
    \renewcommand\thefootnote{\fnsymbol{footnote}}%
    \renewcommand\thempfootnote{\fnsymbol{mpfootnote}}%
    \footnotetext[0]{#1}%
    \egroup
}

\usepackage[autostyle, english = american]{csquotes}
\MakeOuterQuote{"}
\usepackage{hyperref}
\usepackage{cleveref}
\usepackage{caption}
\usepackage[most]{tcolorbox}
\newtcolorbox{infobox}[2][]{
  breakable,
  colback=black!5!white,
  colframe=black!75!white,
  fonttitle=\bfseries,
  title={#2},
  #1
}
\usepackage{tabularx}

\AtBeginDocument{ 
    \fancypagestyle{firstpagestyle}{
        \fancyhf{}
        \fancyfoot[L]{\sffamily\footnotesize \workshopname}%
        \fancyfoot[C]{\sffamily\footnotesize \thepage}
    }
    \fancyhf{}
    \fancyhead[L]{\sffamily\footnotesize\shorttitle}
    \fancyhead[R]{\sffamily\footnotesize\shortauthors}
    \fancyfoot[L]{\sffamily\footnotesize\workshopname}%
    \fancyfoot[C]{\sffamily\footnotesize\thepage}
    \extrafootertext{}
}

\usepackage{lipsum} 

\usepackage{xr}

\begin{document}

\title[Code in Space]{Code in Space: How Multimodal Human-AI Experience Can Reshape the Future of Tech Creation}

\author{Ilya Zakharov}
\email{ilia.zaharov@jetbrains.com}
\affiliation{%
  \institution{JetBrains Research}
  \city{Belgrade}
  \country{Serbia}
}

\author{Ekaterina Koshchenko}
\email{ekaterina.koshchenko@jetbrains.com}
\affiliation{%
  \institution{JetBrains Research}
  \city{Amsterdam}
  \country{Netherlands}
}

\author{Agnia Sergeyuk}
\email{agnia.sergeyuk@jetbrains.com}
\affiliation{%
    \institution{JetBrains Research}
    \city{Belgrade}
    \country{Serbia}
}

\begin{abstract}
Advances in artificial intelligence (AI) continue to reshape digital product development, yet the day-to-day tools for developers and designers remain bound to flat screens and 2D inputs. The intersection of AI and Extended Reality (AI-XR) introduces powerful multimodal interaction channels, such as gaze, motion, or spatial computing, that can enrich the existing Human-AI experience. A critical challenge for utilizing this multimodal opportunities lies in the understanding of how to combine these elements into a cohesive, high-level creative environment. Our study maps this territory through a thematic analysis of semi-structured interviews with 13 AI-XR experts. Categorizing over 150 topics through thematic analysis, we outline five core dimensions of this evolving landscape: professional creation, AI as a contextual layer, new interaction paradigms, adoption frictions, and ethics and human position. Our analysis reveals that besides the critical hardware constraints, the future of AI-XR for tech creation is dependent on addressing human cognitive limits. Ultimately, succeeding in the new multimodal Human-AI experience paradigm requires moving past flat-screen metaphors to design new types of interactions that selectively manage human attention while protecting user agency.

\end{abstract}


\maketitle

\section{Introduction}

The landscape of tech creation (including software engineering, user interface and user experience (UI/UX) design, and data science) is rapidly evolving, driven by advances in artificial intelligence (AI). AI is already reshaping development workflows, from code generation and debugging to design prototyping, and its impact on software development is a major area of research \cite{houck_space_2025}. However, while technology itself is rapidly advancing, the modes of human-computer interaction (HCI) have evolved much more slowly, still relying primarily on manual keyboard and mouse inputs confined to 2D screens. 

Extended Reality (XR) offers a path to go beyond the manual input and flat-screen limitations. Research into XR for software development, while less mature than AI, has already demonstrated innovative ways to visualize and interact with complex systems. Projects like ExplorViz \cite{hasselbring_explorviz_2020} and CodeCity XR \cite{moreno-lumbreras_codecity_2023} use immersive 3D environments to improve spatial comprehension of code structures, while others explore embodied interaction for programming education \cite{hedlund_blocklyvr_2023, vincur_cubely_2017}.

Recent progress in AI and XR technologies has begun to merge these ideas, enabling intelligent interaction environments that adapt to users and their context. This convergence revives earlier concepts such as "intelligent virtual environments"~\cite{luck_applying_2000}. This concept has been coined for more than two decades, but is now becoming technologically feasible due to the modern intersection between XR hard- and software and sophisticated AI models (sometimes referred to AIxVR or AI-XR, \emph{e.g.,}~\cite{hirzle_when_2023, qayyum_secure_2024}). In 2023, the use AI to support interaction in XR constituted 15\% of all research in the field of AI-XR~\cite{hirzle_when_2023}. Today, this question is starting to become even more important \cite{killough_xr_2025}. XR hardware together with AI tooling is uniquely positioned to meet the demand for the new ways of interaction for tech creation, offering an unprecedented richness of multimodal interaction modes within a single, accessible platform. 

In our research, we want to understand how experts in this emerging field see the pathways for the future development of multimodal AI-XR interaction modes that can support tech creation. By \emph{‘multimodal’} here we mean combining more than two input channels and/or coordinated input–output of the tool. By \emph{‘tools for tech creation’} here we understand tools that support tasks across the tech-creation lifecycle, especially software engineering (code writing, comprehension, debugging, review) but also hardware/UI-UX/data science work.

Broadly, our research question can be formulated as follows: \textbf{\textit{what can be the emerging and new ways of interaction in AI-XR that can enhance the experience of tech creators during their work}}? More specifically, we are interested in understanding how experts in XR and, particularly, the intersection between AI and XR fields think about:
\begin{itemize}
    \item \textbf{(RQ1)} What are the needs and recurring pain points of tech creators that can be addressed with multimodal AI-XR tools? 
    \item \textbf{(RQ2)} What types of multimodal interaction techniques are the most promising in addressing these needs and what the AI-XR tooling possibilities for tech creation could look like?
    \item \textbf{(RQ3)} What are the main constraints for building multimodal AI-XR tools for tech creation today?  
\end{itemize}

\section{Background}

The intersection of AI and XR is creating new possibilities for Human-AI Experience (HAX), particularly in the context of tech creation. While this convergence is still in its early stages, growing professional and research interest suggests that immersive and multimodal environments could significantly reshape how people design, develop, and interact with technology. 

\subsection{AI in tech creation}

The widespread adoption of the AI-powered code assistants and coding agents is already fundamentally altering modern software engineering practices. Beyond simple autocompletion, LLMs are now capable of generating entire functions, classes, and even simple applications from natural language prompts \cite{zhang_survey_2024}. 

Research has already demonstrated objective productivity gains: a 2023 study on GitHub Copilot found that developers completed programming tasks significantly faster (over 55\%) when using the AI assistant \cite{peng_impact_2023}. Another large-scale field experiment with nearly 5,000 professional developers across three enterprises showed that granting access to an AI coding assistant increased weekly task completion by 26\%, with larger gains for newer or more junior engineers (evidence from real deployments rather than lab tasks, \cite{cui_effects_2024, houck_space_2025}).

AI's role is also expanding across the entire software development lifecycle. In software testing, LLMs are being leveraged to automatically generate unit tests: recent studies demonstrate their effectiveness in achieving high code coverage, \cite{zapkus_unit_2024}, although a considerable proportion of the generated tests might be incorrect and still require human overview \cite{bhatia_unit_2024}. For debugging and maintenance, advanced models are now capable of automated program repair, suggesting fixes for complex bugs by analyzing the code's context \cite{anand_comprehensive_2024}. This extends to code comprehension and documentation, where AI tools can summarize complex codebases or automatically update comments to reflect code changes, thereby reducing the cognitive load on developers \cite{dvivedi_comparative_2024}.

While transformative, AI advancements are still predominantly delivered through traditional 2D interfaces. This reliance on conventional HCI paradigms limits the potential for more intuitive, multimodal, and spatially-aware interactions.

\subsection{XR and tech creation}

XR technologies offer features well-suited for complex cognitive tasks. They can enhance immersion and attention, enabling deeper concentration \cite{liu_reality_2025} and supporting higher experiences of flow and creativity compared to 2D video conferencing \cite{macchi_virtual_2024}. Studies show that XR can improve teamwork \cite{aufegger_virtual_2022} and even lower perceived cognitive load \cite{chiossi_designing_2025}. While practical adoption has been historically hindered by technological limitations and risks like simulator sickness \cite{souchet_design_2023}, rapid hardware advancements in display fidelity, tracking, and ergonomics are mitigating these issues \cite{kemeny_getting_2024}, making XR an increasingly viable platform for professional tech creation. While numerous studies have already examined AI’s impact on tech creation, much less research addresses how XR technologies may affect the practice of creating new technological solutions. 

Recent developments in this area have enabled innovative ways to visualize, interact with, and collaboratively understand complex software systems. For instance, projects such as Primitive \cite{noauthor_primitive_2025}, ExplorViz \cite{hasselbring_explorviz_2020}, and CodeCity XR \cite{moreno-lumbreras_codecity_2023} utilize immersive 3D environments to enhance spatial comprehension of code structures. Primitive, for example, converts traditional source code into interactive 3D structures, allowing distributed teams to collaboratively explore codebases through intuitive, spatial navigation. Similarly, ExplorViz and CodeCity XR utilize a "software city" metaphor in XR, enabling developers to physically navigate and analyze software architecture, significantly enhancing understanding of complex structures compared to traditional desktop interfaces. ExplorViz’s multi-user modes are also designed to enable collaborative software visualizations, promising to improve program comprehension. Beyond visualization, several XR projects emphasize embodied interaction and educational potential in coding environments. BlocklyXR \cite{hedlund_blocklyvr_2023} and Cubely \cite{vincur_cubely_2017} integrate physical movement and tangible manipulation into programming exercises (programming puzzles in Cubely) or simple tasks (creating and debugging functions in BlocklyXR), showing that bodily engagement can increase motivation and enjoyment without compromising accuracy. Similarly, projects like Hack.VR \cite{kao_hackvr_2020} and Imikode \cite{sunday_usability_2023} are moving coding education into immersive, game-like environments. While end-to-end coding isn’t yet ready for XR, simple workflows are already effective. 

Evaluations \cite{kao_hackvr_2020, sunday_usability_2023, hedlund_blocklyvr_2023} of these XR educational environments indicate strong user enthusiasm and improved motivation, suggesting XR's potential as a compelling medium for programming education, though formal assessments of learning gains remain preliminary. 

\subsection{AI-XR intersection for tech creation}

While the convergence between AI and XR is still emerging, tech creation might appear one of its most prominent application areas. Currently, the most notable topic for study in this area is using AI to build and interact with 3D/spatial content within an immersive environment. For example, the Deep3DVRSketch project \cite{chen_rapid_2024} turns VR sketches from novices into high‑quality 3D models via diffusion models. LLMR framework \cite{torre_llmr_2024} gives the creators an opportunity to prompt mixed‑reality worlds in real time. Early usability studies of these tools show that users report faster modeling than conventional methods \cite{chen_rapid_2024, torre_llmr_2024}. One more interesting emerging area of AI-XR research lies in increasing AI understandability (e.g., by visualizing neural networks in VR, \cite{leitao_computational_2020}), and explainability (for example, by providing immersive interfaces to train machine learning (ML) models for non-experts, \cite{hilton_interactml_2021}). 

However, a primary research challenge is the lack of tools. To bridge this gap, frameworks are being developed to accelerate human-centered AI-XR innovation. A distinct line of work has begun to focus specifically on tech creation itself as the target application of AI-XR, rather than treating coding as an incidental use case. González-Barahona's \cite{gonzalez-barahona_software_2024} vision paper anticipated how IDEs would be reshaped when generative AI and XR co-evolve as the dominant development interface, and the emergence of dedicated venues such as the VARSE workshop at ASE 2025 \cite{Wang2025Second} signals that VR/AR software engineering is now consolidating into its own research community. On the tooling side, DreamCodeVR \cite{Giunchietal2024} uses LLMs to translate spoken language into executable behavior code inside a live VR application, letting non-programmers author object behaviors by voice, while Brown and Mulder \cite{brown2025large} integrate an LLM agent into an immersive binary reverse-engineering environment where it queries analysis tools and generates 3D call-graph visualizations aligned with the analyst's current task, though they report that output quality still varies widely. To lower the engineering barrier for this kind of prototype, XARP \cite{caetano2026xarp} exposes a Unity XR client to Python (and to AI agents via the Model Context Protocol), and reports that AI agents using it consume 19\% fewer tokens than when writing equivalent C\# Unity code; the more recent Vibe Coding XR extends the XR Blocks  \cite{li_xr_2025} direction toward intent-driven "vibe coding" for immersive experiences. Looking further out, Gu and colleagues \cite{gu2025promises} reframe AI-powered XR glasses as a new class of \textit{embodied software} and flag the resulting open problems — validation of spatial capabilities, explainability, security and privacy — as core research challenges at the AI–SE–XR intersection.

The theoretical promise of immersive development is already resonating strongly within the professional community. In the 2024 JetBrains Developer Ecosystem Survey \cite{developer_ecosystem_survey_software_2024}, 49\% of respondents claimed that they \textit{“would love to try”} a virtual reality headset for coding.  XR is unique for tech creation because it offers new input (e.g., gaze tracking, motion tracking) and output (immersive visualization, spatial interaction) modalities in one place and without complicated lab equipment, making it invaluable as a source for new interaction modes for emerging AI tools. In the future, a software engineer debugging a distributed system could simply ask AI-XR tool, "Where is the latency bottleneck?" and have the entire system architecture materialize around them, with the problematic node glowing red. Our research goal is to understand how the human-computer interaction during tech creation can be enhanced with the progress in multimodal AI-XR tools to make this future closer.

\section{Methodology} 

To address our research questions, we adopted a qualitative, exploratory approach aimed at capturing expert perspectives on the emerging field of multimodal AI–XR for tech creation. Specifically, we conducted semi-structured interviews with domain experts to identify key needs, opportunities, and constraints shaping this space. The semi-structured interview questions are summarized below in "Interview Script" box in the Section~\ref{box:script} below. 

We used a \textbf{design-space approach}, in the sense of Heape \cite{heape_design_2007}: the design space as a metaphor for the field of possibilities a designer might consider when meeting a brief. The approach has been productive for organizing emerging technologies \cite{turmo_vidal_design_2021, beyeler_bionic_2025, sergeyuk_bridging_2025} and has recently been applied to XR specifically \cite{davari_towards_2024}. We gathered the qualitative material through \textbf{semi-structured interviews with experts at the intersection of AI and XR}. We analyzed those interviews using thematic analysis following  Braun and Clarke's six-phase approach \cite{braun_using_2006}: 1) Familiarization with data; 2) Generating initial codes; 3) Searching for themes; 4) Reviewing themes; 5) Defining and naming themes; and 6) Writing the report.

Expert recruitment proceeded along two strata. For academic experts, we identified individuals serving in program-committee or editorial roles at top venues from the past five years (using the Hirzle et al., 2023 scoping review \cite{hirzle_when_2023} as a starting point, updated against 2025 Google Scholar metrics, the latest interviews were conducted in February 2026), as well as recent keynote speakers and award recipients. For industry experts, we identified members of AI-XR working groups, maintainers of widely used open-source repositories in the space, patent holders, and individuals in senior product roles (i.e. Principal, Staff, Director, Head) at relevant companies.

Sample size was determined using the \textbf{information-power approach} of Malterud and colleagues \cite{malterud_sample_2016}, considering aim breadth, sample specificity, theory position, dialogue quality, and analysis strategy. Anticipating a broad aim, a specific sample, an absence of established theory, high-quality interviews, and primarily cross-case analysis yielded moderate information power and an initial target of 12–20 participants, split evenly across academic and industry experts.

We also operationalized \textbf{thematic saturation} (following Francis with colleagues \cite{francis_what_2010}): an initial pass at 8 participants, with the analysis updated for every additional three, and saturation declared when three consecutive interviews introduced no new parent themes. We began with \textbf{two pilot interviews}, after which we refined the question set. Most notably, we tightened the section on interaction concepts, which had been too broad. From there we conducted eleven further interviews, for a total of thirteen, with experts from institutions including Cambridge, Aarhus, Stuttgart, and Meta. The thematic analysis coding itself was conducted by two of the authors working independently for the first 8 interviews (Phases 1–4), who then reconciled any disagreement through discussion until a consensus was reached. The remaining interview was coded by only one of the coders and verified by the other coder. 

The study was conducted in accordance with JetBrains ethical standards, adhering to the values and guidelines outlined in \cite{iccesomar_international_2025}. Each interview lasted 45-90 minutes, with an additional ~2 hours per participant for preparation, scheduling, follow-ups, and transcript verification, yielding an estimated 3-3.5 researcher-hours per participant. Overall, it took us 4 months to collect data from the first to the last interview.

\begin{infobox}[unbreakable]{Interview Script} \label{box:script}

\textbf{Introduction to the interview: }

\begin{itemize}
    \item Could you briefly describe your journey with AI-XR field? What kind of projects have you participated in? 
    \item What kind of problems are you trying to solve or see as critical to solve in the AI-XR space?
\end{itemize}

\textbf{The most promising AI-XR prototypes and interaction techniques in general:}
\begin{itemize}
    \item Which current prototypes in the AI-XR field do you find most innovative or instructive and why?
    \item Where does AI most clearly improve XR experiences?
    \item Where does XR most clearly improve how people use or understand AI?
    \item Are there any new research that introduces any promising interaction techniques? 
    \item Which multimodal inputs in XR matter most? 
    \item Which multimodal outputs help most?
    \item What should not be done with AI-XR? Which direction of AI-XR development is not promising?
\end{itemize}

\textbf{Multimodal interaction for tech creation}
\begin{itemize}
    \item Can XR help with your tech-creation problems today?
    \item What XR tools are you expecting/anticipating for to improve tech creation?
    \item What kind of AI products for tech creation in XR do you expect to appear?
    \item In your opinion, what would be an ideal way of interacting with such tools?
    \item What must be solved to build them?
    \item Are there any emerging AI-XR tools or prototypes for tech creation that look promising? 
\end{itemize}

\textbf{Final words:}
\begin{itemize}
    \item Is there anything we haven't discussed that you think is crucial for designing the future AI-XR tools for tech creation?
\end{itemize}
\end{infobox}

\section{Results}
\subsection{Emerging themes}

The final codebook yielded more than a hundred fifty unique codes. We grouped all codes into the following five overarching themes: 1) AI-XR for creation and professional work; 2) AI as a contextual layer for XR; 3) The new interactions paradigms; 4) Technology readiness and adoption friction; 5) Ethics, control, and the human position. These themes and the codes are displayed in Table~\ref{tab:codebook}.

\begin{table*}[!htbp]
  \caption{Overview of the final codebook: five overarching themes, their guiding questions, and second-level subgroups (156 codes in total).}
  \label{tab:codebook}
  \small
  \begin{tabularx}{\textwidth}{@{}p{0.24\textwidth}p{0.22\textwidth}Xc@{}}
    \toprule
    \textbf{Theme} & \textbf{Guiding question} & \textbf{Subgroups (number of codes)} & \textbf{Total} \\
    \midrule
    AI-XR for creation, collaboration, and professional work
      & How does AI-XR change how people make things?
      & Spatial workspaces and portable offices (6); Spatial visualization and embodied engineering (5); AI-assisted prototyping and design (9); Collaboration and representation (2); Knowledge work and the changing creator role (6); Training and immersive professional domains (3)
      & 31 \\
    \addlinespace
    AI as contextual intelligence layer
      & Can AI make XR environments understand and adapt to you?
      & World models and scene understanding (5); Human signals and behavioral context (4); Intent inference and mediation (7); Personalization, attention, and proactivity (8); Situated augmentation and environment control (9)
      & 33 \\
    \addlinespace
    Readiness, adoption, and ecosystem friction
      & What is actually blocking XR-AI from becoming mainstream?
      & Hardware and infrastructure bottlenecks (7); Sensing, precision, and system performance (3); Wearability, ergonomics, and public acceptability (9); Adoption value and workflow friction (9); Paradigm and ecosystem immaturity (10)
      & 38 \\
    \addlinespace
        New interaction paradigms
      & How do humans communicate intent to AI-XR systems?
      & Interaction paradigm shift (7); Multimodal intent orchestration (8); Input channels and wearable controls (7); Embodied presence and multimodal feedback (8); Interaction breakdowns (4)
      & 34 \\
    \addlinespace
    Ethics, control, and the human position
      & As AI-XR systems grow more capable and ambient, what do humans lose or risk?
      & Privacy, surveillance, and security (5); Regulation, governance, and social norms (5); Human agency and dependence (3); Transparency, accountability, and ownership (3); Manipulation, commercialization, and wellbeing (4)
      & 20 \\
    \midrule
    \multicolumn{3}{r}{\textbf{Total codes}} & \textbf{156} \\
    \bottomrule
  \end{tabularx}
\end{table*}

\textbf{AI-XR for creation and professional work.} This is the cluster most directly related to our main question: what is the future of technology creation. Here the main arising questions are centered around productive use: XR environment as a place where knowledge workers, designers, and engineers actually build things. The emerging interaction insight is that the boundary between authoring, coding, and designing is dissolving in XR, and AI is accelerating that dissolution. One unexpected topic that stood out here and genuinely surprised us was framing the XR for robotics as a potential sandbox for training the next generation of machine learning models. The idea is the following: if you put simulated robots into highly controlled virtual environment with rich structured multimodal data, and then create an interaction loop within this virtual environment, you then can use that loop to generate the kind of physical interaction data that today’s language and vision models mostly lack.

Participant 1: \textit{Idea of reality proxy where you kind of use semantic mapping through AI of all the physical environment around you and you semantically map both the objects and their environment around you. This product integration allows you to control drones. It allows you to look into buildings if you have a 3D model of a building}.

\textbf{AI as a contextual layer}. The core idea across this second cluster is based on the fact that XR generates enormous contextual data (gaze, location, body, task state), and AI is what makes that data actionable in real time. According to our experts, the most promising direction today is personalized multimodality that decodes implicit intent. Combining gaze, gestures, voice, and context might be used infer what a person is about to do, not just execute what they explicitly say. Another important topic of discussion here was that the hard bottleneck is shifting from technology to cognition. Building the contextual layer is already becoming technically possible, but the hard thing is to shape it according to the human needs. Hardware will keep improving, but human attention is finite. Several experts argued that the cognitive ceiling will soon become more limiting than the hardware ceiling, and the AI-based context may become as important for creating diminished reality (selectively removing information) as for augmented reality. 

Participant 1: \textit{XR is the kind of interface for us to interact with AIs and AIs will be more grounded in the human experiences with all these new smarter AR glasses coming out}. 

\textbf{New interactions paradigms.} This third cluster is the most populated and is related to AI as a contextual layer. With the rich context available and new AI-based systems and tools becoming more widely adopted, traditional user experience might not be longer sufficient, demanding reinvention of interaction scenarios. The tension here is between expressiveness of new inputs (i.e. you can gesture, look, speak, move with varying amount of information with different inputs) and reliability needed for new paradigms (i.e. the system misunderstands, fatigues you, or fires unintentionally). The 2D desktop paradigm cannot be just ported in 3D, something new has to be invented. Repeating windows, mice, and folders in 3D is the failure mode. The field is still searching for its "desktop metaphor" equivalent for spatial computing.

\textbf{Adoption frictions.} This fourth cluster captures both technical (e.g. hardware, latency, precision of existing solutions) and social/cultural debt (e.g. XR is still weird to wear, awkward in public, not proven valuable enough to justify switching). Any new viable multimodal interaction paradigm have to address these frictions that people experience in order to succeed. 

Participant 11: \textit{The current state of hardware is under the desired level… in terms of batteries, in terms of computational power, in terms of optics}.

And it also captures social/cultural debt (e.g. XR is still weird to wear, awkward in public, not proven valuable enough to justify switching).

Participant 11: \textit{You can’t imagine a bunch of people walking in the streets in headsets like Quest or Apple Vision Pro. So it’s crazy. It’s not socially appropriate. It’s not convenient in many ways}.

\textbf{Ethics, control, and the human position.} The fifth cluster is related to the human consequences of more widespread use of AI-XR systems. Because XR and AI rely on sensing bodies, environments, behavior, and attention, they raise concerns about privacy, agency, social norms, regulation, authorship, and well-being. The essence is that the future of any new future AI-XR interaction paradigm depends not only on technical possibility, but also on whether these systems remain acceptable, safe, and humane.

Participant 1: \textit{If we all walk around with two front-facing cameras and a couple microphones and they are always on for live AI integration... I would prefer there was a lot of testing and regulations in place before that was realized but it is super promising as a technology.}

\subsection{Answering research questions}

We also had the specific research questions we have chosen for the current study. We repeat them below with the relevant study results.

\textbf{RQ1: What are the needs and recurring pain points of tech creators that multimodal AI-XR tools could plausibly address?} 

A recurring theme was that the real bottleneck is cognitive support. The interviews suggest a broad problem: creators struggle to move between ideas, code, spatial artifacts, complex systems, and collaborators using tools that are still mostly screen-based, text-based, and fragmented. Multimodal AI-XR tools could plausibly address these pain points by combining AI generation, spatial visualization, voice, gesture, gaze, and shared immersive environments. AI could help \textbf{generate, adapt, or refine 3D content} from natural language, sketches, gestures, references, or environmental context.

Participant 9: \textit{This is a new medium. Medium requires content. Content creation for multimedia and 3D experiences is notoriously expensive as you can tell from the budget of of games. AI is the way out to manage the content creation}. 

XR could let creators inspect and modify these outputs spatially instead of only through 2D tools. XR could \textbf{make complex systems visible as spatial structures}. XR could \textbf{provide an extended workspace} where code, diagrams, documentation, simulations, and AI assistants are arranged around the user. 

Participant 2: \textit{I think it's a great idea to have a portable office that is even better or more versatile than an office stationary setting}.

AI could organize this space based on task context and personal workflow. Finally, XR could \textbf{create shared workspaces with avatars}, shared models, annotations, and spatial references, while AI could help \textbf{decode intent, summarize discussions, translate ideas into prototypes}, or mediate communication between collaborators. 

Participant 7: \textit{AI can get me to that prototyping stage quicker so I don't have to spend that much time on developing things that are trivial, do not matter for my research and they have been done before}.

Two distinctive needs surfaced more often than we expected. The first was \textbf{embodied telepresence}: realistic, AI-driven avatars with the possibility of richer remote collaboration, which several participants characterized as a potential "killer feature" for distributed engineering teams. The second was \textbf{always-on contextual awareness}: cameras and sensors that continuously model the user's environment so the AI can act on shared, grounded context rather than re-asking what is around it. 

Participant 1: \textit{For AI to be really useful the camera should be… always on basically so that the AI can get the context right: what are you doing and what’s going on; the connection should be always on}. 

\textbf{RQ2: Which multimodal interaction techniques are most promising for addressing those needs, and what might the resulting tooling look like?} 

The most promising multimodal interaction techniques are those that combine explicit user commands with implicit contextual signals. Rather than transplanting desktop GUIs into XR, future tools should let tech creators interact through combinations of voice, gaze, gesture, hand movement, spatial pointing, sketching, and environmental sensing for \textbf{personalized multimodality}. The key value is not any single modality, but the AI’s ability to interpret them together, personalize them to the user and task, and turn vague spatial intentions into precise actions. 

Participant 1: \textit{If you can decode the user's intent, you can also place things that are interesting to the user. So for programming or for general modeling you can always prepare the set of tools that you expect the user to need}.

The resulting tooling would likely take the form of generative 360-degree workspaces, scan-to-prototype pipelines, spatial IDEs, and collaborative AI-XR environments that bridge physical and digital creation. Here, AI can \textbf{arrange spatial layouts that dynamically rearrange information}, generate prototypes, and bridge physical and digital surfaces. Both directions presume a closer integration with AI assistants than with classic IDEs. 

Participant 3:\textit{The idea of application… may just disappear because the operating system now can provide features that the user need at the moment when they need… application developers… are not going to work on applications but rather skills}.

\textbf{RQ3: What are the main constraints — technical, cognitive, organizational — for building such tools today?} 

Four constraints were unanimous across the sample: (1) hardware form factor, (2) AI reliability and domain knowledge, (3) paradigm immaturity, and (4) regulation and privacy. Talking about XR hardware form factor, headsets today are still too heavy, batteries too short-lived, and ergonomics insufficient for sustained work. A "few more device generations" is the typical estimate for when hardware will be ready for mass-market rather than XR enthusiasts. The biggest AI constraint was reliability and domain knowledge. General-purpose models do not yet have enough XR-specific or developer-specific grounding to be trustworthy in this setting; hallucination rates remain a blocker. Paradigm immaturity was the biggest human-centered concern. The field is still waiting for a natural interaction metaphor, its equivalent of the "desktop", and no candidate has yet emerged as obviously right. 

Participant 2: \textit{We have to fundamentally rethink how we interact with these tools. We have to learn how to share control between a human and a machine... You had to teach people how to use a mouse and you're going to have the same when you have this innovative 3D interfaces. We need to think about what are some good metaphors that can help people understand this}.

Finally, talking about \textbf{regulation and privacy:} continuous capture of gaze, environment, and physiology raises substantial legal and ethical issues that the industry has not yet resolved. Even if the technology works, organizations may hesitate to adopt it because of privacy, compliance, cost, unclear productivity gains, and lack of interoperability with existing tools.

\section{Discussion}

Our findings suggest that the trajectory of multimodal HAX for tech creation will be determined less by what becomes technically possible and more by what remains cognitively sustainable. Across all five themes, experts kept returning to the limits of human attention: expressive multimodal input is valuable if it does not fatigue or misfire, contextual AI is the most useful if it filters relevant information, and spatial workspaces are only productive if they organize information instead of scattering it. This reframes the design goal of the field. Where earlier work on immersive development environments largely asked how much more we can show developers, e.g., through software-city visualizations \cite{hasselbring_explorviz_2020, moreno-lumbreras_codecity_2023}, our participants effectively asked how much less the system can demand of them. The most promising interaction techniques that we identified were combinations of explicit commands (voice, gesture, pointing) with implicitly sensed context (gaze, task state, environment), mirroring the direction of recent prototypes such as gaze-grounded embodied agents \cite{bovo_embardiment_2025}, and suggest that the field's "desktop metaphor moment" will likely come from attention-aware orchestration of modalities rather than from any single new input channel. Notably, the experts' vision converges on AI as the integrative layer that makes multimodality coherent: the value proposition of XR for tech creation is inseparable from AI's ability to interpret, personalize, and selectively surface context.

Our results complicate the straightforward "coding in the headset" narrative that motivates much of the popular interest in this space \cite{developer_ecosystem_survey_software_2024}. Some of the experts we have interviewed were skeptical about moving routine code work into them, preferring hybrid setups in which 2D screens remain the precision instrument and XR adds value where spatial information might actually be informative: system architecture, debugging distributed behavior, collaborative review, and prototyping. This implies that near-term AI-XR tooling for tech creation should be conceived as an extension of, not a replacement for, existing development environments. It also foregrounds a tension: the same always-on sensing that makes contextual AI powerful (continuous capture of gaze, environment, and behavior) is precisely what raises the most serious concerns about privacy, agency, and social acceptability. Designing for selective attention management while protecting user agency is therefore not only a usability requirement, but an ethical one. A widespread adoption of these technologies will likely hinge on whether these two demands can be satisfied simultaneously.

\section{Future Work}

As an exploratory interview study with 13 experts, our findings map the design space rather than validate it; the sample, while information-rich, skews toward academic and large-industry perspectives, and expert foresight is an imperfect predictor of actual practice. Future work should therefore triangulate this expert-derived map along three lines. First, the needs and pain points identified under RQ1 should be \textbf{validated with practicing tech creators} without XR expertise to test whether the cognitive bottlenecks experts anticipate match those practitioners experience. Second, the interaction concepts surfaced under RQ2 (personalized multimodality, attention-aware spatial layouts, hybrid 2D/3D workflows) should be \textbf{embodied in low- and mid-fidelity prototypes} and evaluated empirically, with particular attention to precision–expressiveness trade-offs and longitudinal fatigue rather than one-session novelty effects. Third, the constraints identified under RQ3 suggest several concrete research programs, such as, for example, \textbf{benchmarking AI reliability on XR- and developer-specific grounding or developing privacy-preserving architectures} for always-on contextual sensing before, rather than after, such systems reach users. We intend to use the resulting design space as scaffolding for this prototype-driven phase, moving from how experts imagine multimodal AI-XR tech creation toward evidence about how creators can actually work within it.

\section{Limitations}

Several limitations qualify our findings. First, our results reflect \textbf{expert foresight rather than observed practice}. Interviews capture what knowledgeable people anticipate, and the history of both AI and XR shows that expert predictions in fast-moving fields are frequently wrong in timing and sometimes in direction. Our design space should therefore be read as a map of plausible pathways, not a forecast. Relatedly, the field itself is evolving quickly: our interviews were conducted over a four-month window ending in February 2026, and the rapid pace of both AI model capabilities and XR hardware releases means some constraints participants described (e.g., latency, battery life, model grounding) may shift substantially even in the short term.

Second, there are \textbf{sampling limitations}. Although our two-strata recruitment strategy and the information-power approach \cite{malterud_sample_2016} were designed to maximize the relevance of a small sample, thirteen experts cannot represent the full breadth of the AI-XR community. Our sample skews toward academic researchers and senior members of large organizations; independent developers, startup practitioners, and — importantly — tech creators without XR expertise are absent. The perspectives of the intended end users of these tools are thus mediated entirely through experts' assumptions about them. Geographic and cultural coverage is likewise limited, which matters for themes such as social acceptability and regulation, where norms differ considerably across regions.

Third, there are \textbf{analytical and positional limitations}. Thematic analysis is interpretive by nature; although two researchers coded the first eight interviews independently and reconciled disagreements through discussion, a different team might have organized the more than 150 codes into different themes, and our saturation criterion (three consecutive interviews with no new parent themes) guards against missing major themes but not minor ones. Our interview script also framed the conversation around tech creation and multimodality, which may have led participants toward these topics more than they would have raised spontaneously. Finally, this research was conducted by an industry research group affiliated with a developer-tools company. While the study followed the ethical guideline and participants had no commercial relationship with us, our institutional position may have shaped both what participants chose to emphasize and our own interpretive lens, particularly around the future of development environments.

\bibliographystyle{ACM-Reference-Format}
\bibliography{references}

@inproceedings{dvivedi_comparative_2024,
	address = {New York, NY, USA},
	series = {{AIware} 2024},
	title = {A {Comparative} {Analysis} of {Large} {Language} {Models} for {Code} {Documentation} {Generation}},
	isbn = {979-8-4007-0685-1},
	url = {https://doi.org/10.1145/3664646.3664765},
	doi = {10.1145/3664646.3664765},
	urldate = {2025-10-31},
	booktitle = {Proceedings of the 1st {ACM} {International} {Conference} on {AI}-{Powered} {Software}},
	publisher = {Association for Computing Machinery},
	author = {Dvivedi, Shubhang Shekhar and Vijay, Vyshnav and Pujari, Sai Leela Rahul and Lodh, Shoumik and Kumar, Dhruv},
	month = jul,
	year = {2024},
	pages = {65--73},
}

@misc{anand_comprehensive_2024,
	title = {A {Comprehensive} {Survey} of {AI}-{Driven} {Advancements} and {Techniques} in {Automated} {Program} {Repair} and {Code} {Generation}},
	url = {http://arxiv.org/abs/2411.07586},
	doi = {10.48550/arXiv.2411.07586},
	urldate = {2025-10-31},
	publisher = {arXiv},
	author = {Anand, Avinash and Gupta, Akshit and Yadav, Nishchay and Bajaj, Shaurya},
	month = nov,
	year = {2024},
	note = {arXiv:2411.07586 [cs]},
}

@article{zapkus_unit_2024,
	title = {Unit test generation using large language models: a systematic literature review},
	issn = {2783-784X},
	shorttitle = {Unit test generation using large language models},
	url = {https://epublications.vu.lt/object/elaba:198450771/},
	doi = {10.15388/LMITT.2024.20},
	language = {eng},
	urldate = {2025-10-31},
	journal = {Lietuvos magistrantų informatikos ir IT tyrimai: konferencijos darbai, 2024 m. gegužės 10 d.},
	author = {Zapkus, Dovydas Marius and Slotkienė, Asta},
	year = {2024},
	note = {Publisher: Vilniaus universiteto leidykla / Vilnius University Press},
	pages = {136--144},
}

@misc{peng_impact_2023,
	title = {The {Impact} of {AI} on {Developer} {Productivity}: {Evidence} from {GitHub} {Copilot}},
	shorttitle = {The {Impact} of {AI} on {Developer} {Productivity}},
	url = {http://arxiv.org/abs/2302.06590},
	doi = {10.48550/arXiv.2302.06590},
	urldate = {2025-10-31},
	publisher = {arXiv},
	author = {Peng, Sida and Kalliamvakou, Eirini and Cihon, Peter and Demirer, Mert},
	month = feb,
	year = {2023},
	note = {arXiv:2302.06590 [cs]},
}

@misc{zhang_survey_2024,
	title = {A {Survey} on {Large} {Language} {Models} for {Software} {Engineering}},
	url = {http://arxiv.org/abs/2312.15223},
	doi = {10.48550/arXiv.2312.15223},
	urldate = {2025-10-31},
	publisher = {arXiv},
	author = {Zhang, Quanjun and Fang, Chunrong and Xie, Yang and Zhang, Yaxin and Yang, Yun and Sun, Weisong and Yu, Shengcheng and Chen, Zhenyu},
	month = sep,
	year = {2024},
	note = {arXiv:2312.15223 [cs]},
}

@misc{torre_llmr_2024,
	title = {{LLMR}: {Real}-time {Prompting} of {Interactive} {Worlds} using {Large} {Language} {Models}},
	shorttitle = {{LLMR}},
	url = {http://arxiv.org/abs/2309.12276},
	doi = {10.48550/arXiv.2309.12276},
	urldate = {2025-10-31},
	publisher = {arXiv},
	author = {Torre, Fernanda De La and Fang, Cathy Mengying and Huang, Han and Banburski-Fahey, Andrzej and Fernandez, Judith Amores and Lanier, Jaron},
	month = mar,
	year = {2024},
	note = {arXiv:2309.12276 [cs]},
}

@inproceedings{chen_rapid_2024,
	address = {Seattle, WA, USA},
	title = {Rapid {3D} {Model} {Generation} with {Intuitive} {3D} {Input}},
	copyright = {https://doi.org/10.15223/policy-029},
	isbn = {979-8-3503-5300-6},
	url = {https://ieeexplore.ieee.org/document/10656598/},
	doi = {10.1109/CVPR52733.2024.01193},
	language = {en},
	urldate = {2025-10-31},
	booktitle = {2024 {IEEE}/{CVF} {Conference} on {Computer} {Vision} and {Pattern} {Recognition} ({CVPR})},
	publisher = {IEEE},
	author = {Chen, Tianrun and Ding, Chaotao and Zhang, Shangzhan and Yu, Chunan and Zang, Ying and Li, Zejian and Peng, Sida and Sun, Lingyun},
	month = jun,
	year = {2024},
	pages = {12554--12564},
}

@inproceedings{bhatia_unit_2024,
	address = {New York, NY, USA},
	series = {{LLM4Code} '24},
	title = {Unit {Test} {Generation} using {Generative} {AI} : {A} {Comparative} {Performance} {Analysis} of {Autogeneration} {Tools}},
	isbn = {979-8-4007-0579-3},
	shorttitle = {Unit {Test} {Generation} using {Generative} {AI}},
	url = {https://dl.acm.org/doi/10.1145/3643795.3648396},
	doi = {10.1145/3643795.3648396},
	urldate = {2025-10-31},
	booktitle = {Proceedings of the 1st {International} {Workshop} on {Large} {Language} {Models} for {Code}},
	publisher = {Association for Computing Machinery},
	author = {Bhatia, Shreya and Gandhi, Tarushi and Kumar, Dhruv and Jalote, Pankaj},
	month = sep,
	year = {2024},
	pages = {54--61},
}

@misc{killough_xr_2025,
	title = {{XR} for {All}: {Understanding} {Developers}' {Perspectives} on {Accessibility} {Integration} in {Extended} {Reality}},
	shorttitle = {{XR} for {All}},
	url = {http://arxiv.org/abs/2412.16321},
	doi = {10.48550/arXiv.2412.16321},
	urldate = {2025-10-31},
	publisher = {arXiv},
	author = {Killough, Daniel and Ji, Tiger F. and Zhang, Kexin and Hu, Yaxin and Huang, Yu and Du, Ruofei and Zhao, Yuhang},
	month = aug,
	year = {2025},
	note = {arXiv:2412.16321 [cs]},
}

@misc{developer_ecosystem_survey_software_2024,
	title = {Software {Developers} {Statistics} 2024 - {State} of {Developer} {Ecosystem} {Report}},
	url = {https://www.jetbrains.com/lp/devecosystem-2024},
	language = {en},
	urldate = {2025-10-30},
	journal = {JetBrains: Developer Tools for Professionals and Teams},
	author = {Developer Ecosystem Survey},
	year = {2024},
}

@misc{iccesomar_international_2025,
	title = {International {Code} on {Market}, {Opinion} and {Social} {Research} and {Data} {Analytics}},
	url = {https://iccwbo.org/news-publications/policies-reports/iccesomar-international-code-market-opinion-social-research-data-analytics/},
	language = {en-US},
	urldate = {2025-10-30},
	journal = {ICC - International Chamber of Commerce},
	author = {ICC/ESOMAR},
	month = sep,
	year = {2025},
}

@misc{noauthor_primitive_2025,
	title = {Primitive},
	url = {http://primitive.io/},
	language = {en},
	urldate = {2025-10-30},
	journal = {Primitive},
	year = {2025},
}

@article{qayyum_secure_2024,
	title = {Secure and {Trustworthy} {Artificial} {Intelligence}-extended {Reality} ({AI}-{XR}) for {Metaverses}},
	volume = {56},
	issn = {0360-0300},
	url = {https://dl.acm.org/doi/10.1145/3614426},
	doi = {10.1145/3614426},
	number = {7},
	urldate = {2025-10-30},
	journal = {ACM Comput. Surv.},
	author = {Qayyum, Adnan and Butt, Muhammad Atif and Ali, Hassan and Usman, Muhammad and Halabi, Osama and Al-Fuqaha, Ala and Abbasi, Qammer H. and Imran, Muhammad Ali and Qadir, Junaid},
	month = apr,
	year = {2024},
	pages = {170:1--170:38},
}

@article{malterud_sample_2016,
	title = {Sample {Size} in {Qualitative} {Interview} {Studies}: {Guided} by {Information} {Power}},
	volume = {26},
	issn = {1049-7323},
	shorttitle = {Sample {Size} in {Qualitative} {Interview} {Studies}},
	url = {https://doi.org/10.1177/1049732315617444},
	doi = {10.1177/1049732315617444},
	language = {EN},
	number = {13},
	urldate = {2025-10-30},
	journal = {Qualitative Health Research},
	author = {Malterud, Kirsti and Siersma, Volkert Dirk and Guassora, Ann Dorrit},
	month = nov,
	year = {2016},
	note = {Publisher: SAGE Publications Inc},
	pages = {1753--1760},
}

@article{francis_what_2010,
	title = {What is an adequate sample size? {Operationalising} data saturation for theory-based interview studies},
	volume = {25},
	issn = {0887-0446},
	shorttitle = {What is an adequate sample size?},
	url = {https://doi.org/10.1080/08870440903194015},
	doi = {10.1080/08870440903194015},
	number = {10},
	urldate = {2025-10-30},
	journal = {Psychology \& Health},
	author = {Francis, Jill J. and Johnston, Marie and Robertson, Clare and Glidewell, Liz and Entwistle, Vikki and Eccles, Martin P. and Grimshaw, Jeremy M.},
	month = dec,
	year = {2010},
	pmid = {20204937},
	note = {Publisher: Routledge
\_eprint: https://doi.org/10.1080/08870440903194015},
	pages = {1229--1245},
}

@inproceedings{hilton_interactml_2021,
	address = {New York, NY, USA},
	series = {{VRST} '21},
	title = {{InteractML}: {Making} machine learning accessible for creative practitioners working with movement interaction in immersive media},
	isbn = {978-1-4503-9092-7},
	shorttitle = {{InteractML}},
	url = {https://doi.org/10.1145/3489849.3489879},
	doi = {10.1145/3489849.3489879},
	urldate = {2025-10-29},
	booktitle = {Proceedings of the 27th {ACM} {Symposium} on {Virtual} {Reality} {Software} and {Technology}},
	publisher = {Association for Computing Machinery},
	author = {Hilton, Clarice and Plant, Nicola and González Díaz, Carlos and Perry, Phoenix and Gibson, Ruth and Martelli, Bruno and Zbyszynski, Michael and Fiebrink, Rebecca and Gillies, Marco},
	month = dec,
	year = {2021},
	pages = {1--10},
}

@article{leitao_computational_2020,
	title = {Computational imaging during video game playing shows dynamic synchronization of cortical and subcortical networks of emotions},
	volume = {18},
	issn = {1545-7885},
	url = {https://journals.plos.org/plosbiology/article?id=10.1371/journal.pbio.3000900},
	doi = {10.1371/journal.pbio.3000900},
	language = {en},
	number = {11},
	urldate = {2025-10-30},
	journal = {PLOS Biology},
	author = {Leitão, Joana and Meuleman, Ben and Ville, Dimitri Van De and Vuilleumier, Patrik},
	month = nov,
	year = {2020},
	note = {Publisher: Public Library of Science},
	pages = {e3000900},
}

@misc{sergeyuk_bridging_2025,
	title = {Bridging {Developer} {Needs} and {Feasible} {Features} for {AI} {Assistants} in {IDEs}},
	url = {http://arxiv.org/abs/2410.08676},
	doi = {10.48550/arXiv.2410.08676},
	urldate = {2025-10-30},
	publisher = {arXiv},
	author = {Sergeyuk, Agnia and Koshchenko, Ekaterina and Zakharov, Ilya and Bryksin, Timofey and Izadi, Maliheh},
	month = aug,
	year = {2025},
	note = {arXiv:2410.08676 [cs]},
}

@article{braun_using_2006,
	title = {Using thematic analysis in psychology},
	volume = {3},
	issn = {1478-0887},
	url = {https://doi.org/10.1191/1478088706qp063oa},
	doi = {10.1191/1478088706qp063oa},
	number = {2},
	urldate = {2025-10-30},
	journal = {Qualitative Research in Psychology},
	author = {Braun, Virginia and Clarke, Victoria},
	month = jan,
	year = {2006},
	note = {Publisher: Routledge
\_eprint: https://doi.org/10.1191/1478088706qp063oa},
	pages = {77--101},
}

@inproceedings{turmo_vidal_design_2021,
	address = {New York, NY, USA},
	series = {{CHI} '21},
	title = {The {Design} {Space} of {Wearables} for {Sports} and {Fitness} {Practices}},
	isbn = {978-1-4503-8096-6},
	url = {https://doi.org/10.1145/3411764.3445700},
	doi = {10.1145/3411764.3445700},
	urldate = {2025-10-29},
	booktitle = {Proceedings of the 2021 {CHI} {Conference} on {Human} {Factors} in {Computing} {Systems}},
	publisher = {Association for Computing Machinery},
	author = {Turmo Vidal, Laia and Zhu, Hui and Waern, Annika and Márquez Segura, Elena},
	month = may,
	year = {2021},
	pages = {1--14},
}

@misc{beyeler_bionic_2025,
	title = {Bionic {Vision} as {Neuroadaptive} {XR}: {Closed}-{Loop} {Perceptual} {Interfaces} for {Neurotechnology}},
	shorttitle = {Bionic {Vision} as {Neuroadaptive} {XR}},
	url = {http://arxiv.org/abs/2508.05963},
	doi = {10.48550/arXiv.2508.05963},
	urldate = {2025-10-30},
	publisher = {arXiv},
	author = {Beyeler, Michael},
	month = aug,
	year = {2025},
	note = {arXiv:2508.05963 [cs]},
}

@article{heape_design_2007,
	title = {The {Design} {Space}: the design process as the construction, exploration and expansion of a conceptual space},
	issn = {978-87-991686-6-8},
	shorttitle = {The {Design} {Space}},
	author = {Heape, Chris},
	year = {2007},
	note = {Publisher: [s.n.]},
}

@article{luck_applying_2000,
	title = {Applying artificial intelligence to virtual reality: {Intelligent} virtual environments},
	volume = {14},
	issn = {0883-9514},
	shorttitle = {Applying artificial intelligence to virtual reality},
	url = {https://doi.org/10.1080/088395100117142},
	doi = {10.1080/088395100117142},
	number = {1},
	urldate = {2025-10-30},
	journal = {Applied Artificial Intelligence},
	author = {Luck, Michael and Aylett, Ruth},
	month = jan,
	year = {2000},
	note = {Publisher: Taylor \& Francis
\_eprint: https://doi.org/10.1080/088395100117142},
	pages = {3--32},
}

@misc{li_xr_2025,
	title = {{XR} {Blocks}: {Accelerating} {Human}-centered {AI} + {XR} {Innovation}},
	shorttitle = {{XR} {Blocks}},
	url = {http://arxiv.org/abs/2509.25504},
	doi = {10.48550/arXiv.2509.25504},
	urldate = {2025-10-27},
	publisher = {arXiv},
	author = {Li, David and Numan, Nels and Qian, Xun and Chen, Yanhe and Zhou, Zhongyi and Alekseev, Evgenii and Lee, Geonsun and Cooper, Alex and Xia, Min and Chung, Scott and Nelson, Jeremy and Yuan, Xiuxiu and Dias, Jolica and Bettridge, Tim and Hersh, Benjamin and Huynh, Michelle and Piascik, Konrad and Cabello, Ricardo and Kim, David and Du, Ruofei},
	month = sep,
	year = {2025},
	note = {arXiv:2509.25504 [cs]},
}

@incollection{kemeny_getting_2024,
	address = {Cham},
	title = {Getting {Rid} of {Motion} {Sickness}},
	isbn = {978-3-031-45263-5},
	url = {https://doi.org/10.1007/978-3-031-45263-5_5},
	language = {en},
	urldate = {2025-09-24},
	booktitle = {Autonomous {Vehicles} and {Virtual} {Reality}: {The} {New} {Automobile} {Industrial} {Revolution}},
	publisher = {Springer International Publishing},
	author = {Kemeny, Andras},
	editor = {Kemeny, Andras},
	year = {2024},
	doi = {10.1007/978-3-031-45263-5_5},
	pages = {83--98},
}

@article{sunday_usability_2023,
	title = {Usability {Evaluation} of {Imikode} {Virtual} {Reality} {Game} to {Facilitate} {Learning} of {Object}-{Oriented} {Programming}},
	volume = {28},
	issn = {2211-1670},
	url = {https://doi.org/10.1007/s10758-022-09634-6},
	doi = {10.1007/s10758-022-09634-6},
	language = {en},
	number = {4},
	urldate = {2025-09-24},
	journal = {Technology, Knowledge and Learning},
	author = {Sunday, Kissinger and Oyelere, Solomon Sunday and Agbo, Friday Joseph and Aliyu, Muhammad Bello and Balogun, Oluwafemi Samson and Bouali, Nacir},
	month = dec,
	year = {2023},
	pages = {1871--1902},
}

@article{chiossi_designing_2025,
	title = {Designing and {Evaluating} an {Adaptive} {Virtual} {Reality} {System} using {EEG} {Frequencies} to {Balance} {Internal} and {External} {Attention} {States}},
	volume = {196},
	issn = {10715819},
	url = {http://arxiv.org/abs/2311.10447},
	doi = {10.1016/j.ijhcs.2024.103433},
	urldate = {2025-09-24},
	journal = {International Journal of Human-Computer Studies},
	author = {Chiossi, Francesco and Ou, Changkun and Gerhardt, Carolina and Putze, Felix and Mayer, Sven},
	month = feb,
	year = {2025},
	note = {arXiv:2311.10447 [cs]},
	pages = {103433},
}

@article{macchi_virtual_2024,
	title = {Virtual reality, face-to-face, and {2D} video conferencing differently impact fatigue, creativity, flow, and decision-making in workplace dynamics},
	volume = {14},
	copyright = {2024 The Author(s)},
	issn = {2045-2322},
	url = {https://www.nature.com/articles/s41598-024-60942-6},
	doi = {10.1038/s41598-024-60942-6},
	language = {en},
	number = {1},
	urldate = {2025-09-24},
	journal = {Scientific Reports},
	author = {Macchi, Gregorio and De Pisapia, Nicola},
	month = may,
	year = {2024},
	note = {Publisher: Nature Publishing Group},
	pages = {10260},
}

@article{aufegger_virtual_2022,
	title = {Virtual {Reality} and {Productivity} in {Knowledge} {Workers}},
	volume = {3},
	issn = {2673-4192},
	url = {https://www.frontiersin.org/journals/virtual-reality/articles/10.3389/frvir.2022.890700/full},
	doi = {10.3389/frvir.2022.890700},
	language = {English},
	urldate = {2025-09-24},
	journal = {Frontiers in Virtual Reality},
	author = {Aufegger, Lisa and Elliott-Deflo, Natasha},
	month = may,
	year = {2022},
	note = {Publisher: Frontiers},
}

@article{souchet_design_2023,
	title = {Design guidelines for limiting and eliminating virtual reality-induced symptoms and effects at work: a comprehensive, factor-oriented review},
	volume = {14},
	issn = {1664-1078},
	shorttitle = {Design guidelines for limiting and eliminating virtual reality-induced symptoms and effects at work},
	url = {https://www.frontiersin.org/journals/psychology/articles/10.3389/fpsyg.2023.1161932/full},
	doi = {10.3389/fpsyg.2023.1161932},
	language = {English},
	urldate = {2025-09-24},
	journal = {Frontiers in Psychology},
	author = {Souchet, Alexis D. and Lourdeaux, Domitile and Burkhardt, Jean-Marie and Hancock, Peter A.},
	month = jun,
	year = {2023},
	note = {Publisher: Frontiers},
}

@article{hasselbring_explorviz_2020,
	title = {{ExplorViz}: {Research} on software visualization, comprehension and collaboration},
	volume = {6},
	issn = {2665-9638},
	shorttitle = {{ExplorViz}},
	url = {https://www.sciencedirect.com/science/article/pii/S2665963820300257},
	doi = {10.1016/j.simpa.2020.100034},
	urldate = {2025-09-24},
	journal = {Software Impacts},
	author = {Hasselbring, Wilhelm and Krause, Alexander and Zirkelbach, Christian},
	month = nov,
	year = {2020},
	pages = {100034},
}

@article{moreno-lumbreras_codecity_2023,
	title = {{CodeCity}: {A} comparison of on-screen and virtual reality},
	volume = {153},
	issn = {0950-5849},
	shorttitle = {{CodeCity}},
	url = {https://www.sciencedirect.com/science/article/pii/S0950584922001732},
	doi = {10.1016/j.infsof.2022.107064},
	urldate = {2025-09-24},
	journal = {Information and Software Technology},
	author = {Moreno-Lumbreras, David and Minelli, Roberto and Villaverde, Andrea and Gonzalez-Barahona, Jesus M. and Lanza, Michele},
	month = jan,
	year = {2023},
	pages = {107064},
}

@inproceedings{hedlund_blocklyvr_2023,
	address = {New York, NY, USA},
	series = {{MUM} '23},
	title = {{BlocklyVR}: {Exploring} {Block}-based {Programming} in {Virtual} {Reality}},
	isbn = {979-8-4007-0921-0},
	shorttitle = {{BlocklyVR}},
	url = {https://dl.acm.org/doi/10.1145/3626705.3627779},
	doi = {10.1145/3626705.3627779},
	urldate = {2025-09-24},
	booktitle = {Proceedings of the 22nd {International} {Conference} on {Mobile} and {Ubiquitous} {Multimedia}},
	publisher = {Association for Computing Machinery},
	author = {Hedlund, Martin and Jonsson, Adam and Bogdan, Cristian and Meixner, Gerrit and Ekblom Bak, Elin and Matviienko, Andrii},
	month = dec,
	year = {2023},
	pages = {257--269},
}

@inproceedings{vincur_cubely_2017,
	address = {New York, NY, USA},
	series = {{VRST} '17},
	title = {Cubely: virtual reality block-based programming environment},
	isbn = {978-1-4503-5548-3},
	shorttitle = {Cubely},
	url = {https://doi.org/10.1145/3139131.3141785},
	doi = {10.1145/3139131.3141785},
	urldate = {2025-09-24},
	booktitle = {Proceedings of the 23rd {ACM} {Symposium} on {Virtual} {Reality} {Software} and {Technology}},
	publisher = {Association for Computing Machinery},
	author = {Vincur, Juraj and Konopka, Martin and Tvarozek, Jozef and Hoang, Martin and Navrat, Pavol},
	month = nov,
	year = {2017},
	pages = {1--2},
}

@book{kao_hackvr_2020,
	title = {Hack.{VR}: {A} {Programming} {Game} in {Virtual} {Reality}},
	shorttitle = {Hack.{VR}},
	author = {Kao, Dominic and Mousas, Christos and Magana, Alejandra and Harrell, D. and Ratan, Rabindra and Melcer, Edward and Sherrick, Brett and Parsons, Paul and Gusev, Dmitri},
	month = jul,
	year = {2020},
	doi = {10.48550/arXiv.2007.04495},
}

@misc{bovo_embardiment_2025,
	title = {{EmBARDiment}: an {Embodied} {AI} {Agent} for {Productivity} in {XR}},
	shorttitle = {{EmBARDiment}},
	url = {http://arxiv.org/abs/2408.08158},
	doi = {10.48550/arXiv.2408.08158},
	urldate = {2025-09-02},
	publisher = {arXiv},
	author = {Bovo, Riccardo and Abreu, Steven and Ahuja, Karan and Gonzalez, Eric J. and Cheng, Li-Te and Gonzalez-Franco, Mar},
	month = mar,
	year = {2025},
	note = {arXiv:2408.08158 [cs]},
}

@misc{houck_space_2025,
	title = {The {SPACE} of {AI}: {Real}-{World} {Lessons} on {AI}'s {Impact} on {Developers}},
	shorttitle = {The {SPACE} of {AI}},
	url = {http://arxiv.org/abs/2508.00178},
	doi = {10.48550/arXiv.2508.00178},
	urldate = {2025-08-28},
	publisher = {arXiv},
	author = {Houck, Brian and Lowdermilk, Travis and Beyer, Cody and Clarke, Steven and Hanrahan, Ben},
	month = jul,
	year = {2025},
	note = {arXiv:2508.00178 [cs]},
}

@inproceedings{hirzle_when_2023,
	address = {New York, NY, USA},
	series = {{CHI} '23},
	title = {When {XR} and {AI} {Meet} - {A} {Scoping} {Review} on {Extended} {Reality} and {Artificial} {Intelligence}},
	isbn = {978-1-4503-9421-5},
	url = {https://dl.acm.org/doi/10.1145/3544548.3581072},
	doi = {10.1145/3544548.3581072},
	urldate = {2025-08-07},
	booktitle = {Proceedings of the 2023 {CHI} {Conference} on {Human} {Factors} in {Computing} {Systems}},
	publisher = {Association for Computing Machinery},
	author = {Hirzle, Teresa and Müller, Florian and Draxler, Fiona and Schmitz, Martin and Knierim, Pascal and Hornbæk, Kasper},
	month = apr,
	year = {2023},
	pages = {1--45},
}

@misc{liu_reality_2025,
	title = {Reality {Proxy}: {Fluid} {Interactions} with {Real}-{World} {Objects} in {MR} via {Abstract} {Representations}},
	shorttitle = {Reality {Proxy}},
	url = {http://arxiv.org/abs/2507.17248},
	doi = {10.1145/3746059.3747709},
	urldate = {2025-08-07},
	author = {Liu, Xiaoan and Jia, Difan and Liu, Xianhao Carton and Gonzalez-Franco, Mar and Zhu-Tian, Chen},
	month = jul,
	year = {2025},
	note = {arXiv:2507.17248 [cs]},
}

@misc{davari_towards_2024,
	title = {Towards {Context}-{Aware} {Adaptation} in {Extended} {Reality}: {A} {Design} {Space} for {XR} {Interfaces} and an {Adaptive} {Placement} {Strategy}},
	shorttitle = {Towards {Context}-{Aware} {Adaptation} in {Extended} {Reality}},
	url = {http://arxiv.org/abs/2411.02607},
	doi = {10.48550/arXiv.2411.02607},
	urldate = {2025-07-30},
	publisher = {arXiv},
	author = {Davari, Shakiba and Bowman, Doug A.},
	month = nov,
	year = {2024},
	note = {arXiv:2411.02607 [cs]},
}

@misc{cui_effects_2024,
	address = {Rochester, NY},
	type = {{SSRN} {Scholarly} {Paper}},
	title = {The {Effects} of {Generative} {AI} on {High} {Skilled} {Work}: {Evidence} from {Three} {Field} {Experiments} with {Software} {Developers}},
	shorttitle = {The {Effects} of {Generative} {AI} on {High} {Skilled} {Work}},
	url = {https://papers.ssrn.com/abstract=4945566},
	doi = {10.2139/ssrn.4945566},
	language = {en},
	urldate = {2024-09-12},
	author = {Cui, Zheyuan (Kevin) and Demirer, Mert and Jaffe, Sonia and Musolff, Leon and Peng, Sida and Salz, Tobias},
	month = sep,
	year = {2024},
}

@inproceedings{gonzalez-barahona_software_2024,
	title = {Software development in the age of {LLMs} and {XR}},
	url = {http://arxiv.org/abs/2404.09789},
	doi = {10.1145/3643796.3648457},
	urldate = {2024-09-03},
	booktitle = {Proceedings of the 1st {ACM}/{IEEE} {Workshop} on {Integrated} {Development} {Environments}},
	author = {Gonzalez-Barahona, Jesus M.},
	month = apr,
	year = {2024},
	note = {arXiv:2404.09789 [cs]},
	pages = {66--69},
}

@article{Wang2025Second,
  author  = {Wang, Xiaoyin and He, Sen and Hassan, Foyzul},
  title   = {The Second International Workshop on Virtual and Augmented Reality Software Engineering},
  journal = {ACM SIGSOFT Software Engineering Notes},
  year    = {2025}
}

@inproceedings{Giunchietal2024,
  author    = {Giunchi, Daniele and Numan, Nels and Gatti, Elia and Steed, Anthony},
  title     = {DreamCodeVR: Towards Democratizing Behavior Design in Virtual Reality with Speech-Driven Programming},
  booktitle = {Proceedings of the 2024 IEEE Conference on Virtual Reality and 3D User Interfaces (VR)},
  year      = {2024},
  pages     = {579--589},
  publisher = {IEEE}
}

@article{caetano2026xarp,
  author    = {Caetano, Arthur and Kumaran, Radha and Jou, Kelvin and H{\"o}llerer, Tobias and Sra, Misha},
  title     = {XARP Tools: An Extended Reality Platform for Humans and AI Agents},
  journal   = {Proceedings of the ACM on Human-Computer Interaction},
  volume    = {10},
  number    = {EICS},
  year      = {2026},
  doi       = {10.1145/3816762},
  publisher = {ACM New York, NY, USA},
  note      = {arXiv preprint arXiv:2508.04108}
}

@inproceedings{gu2025promises,
  author    = {Gu, Ruizhen and Zhang, Jingqiong and Rojas, Jos{\'e} Miguel and Shin, Donghwan},
  title     = {On the Promises and Challenges of AI-Powered XR Glasses as Embodied Software},
  booktitle = {Proceedings of the 2025 International Conference on AI-powered Software (AIware)},
  pages     = {207--212},
  year      = {2025}
}

@inproceedings{brown2025large,
  author    = {Brown, Dennis G. and Mulder, Samuel},
  title     = {Large Language Models as Visualization Agents for Immersive Binary Reverse Engineering},
  booktitle = {Proceedings of the 2025 IEEE Working Conference on Software Visualization (VISSOFT)},
  year      = {2025}
}

\end{document}